\documentclass[preprint,12pt]{elsarticle}
\usepackage{graphicx}
\usepackage{braket}
\usepackage{amssymb}
\usepackage{amsmath}
\usepackage[utf8]{inputenc}
\usepackage{bm}
\usepackage{xcolor}
\usepackage{hyperref}
\journal{Physica B: Condensed Matter}

\makeatletter
\def\ps@pprintTitle{%
  \let\@oddhead\@empty
  \let\@evenhead\@empty
  \let\@oddfoot\@empty
  \let\@evenfoot\@empty}
\makeatother

\begin{document}
\begin{frontmatter}
\title{Ferrimagnetic and Haldane-type phases in a mixed-spin $1$-$\tfrac{1}{2}$-$\tfrac{1}{2}$ quantum trimer chain}
\author[ufpe]{A. Felinto}
\author[ufpe]{R. R. Montenegro-Filho\corref{cor1}}
\ead{rene.montenegro@ufpe.br}
\cortext[cor1]{Corresponding author}
\address[ufpe]{Laborat\'{o}rio de F\'{i}sica Te\'{o}rica e Computacional, Departamento de F\'{i}sica, Universidade Federal de Pernambuco, 50760-901 Recife-PE, Brazil}

\begin{abstract}
Bipartite Lieb-Mattis ferrimagnetism and the symmetry-protected
Haldane phase are paradigmatic mechanisms in quasi-one-dimensional
quantum magnets. Both emerge, in distinct regimes, in a mixed-spin
$1$-$\tfrac{1}{2}$-$\tfrac{1}{2}$ Heisenberg trimer chain with
antiferromagnetic backbone exchange $J$ and a side spin-$\tfrac{1}{2}$
coupled to each backbone spin by an exchange $J_t$ of either sign.
Using the density matrix renormalization group, we compute
magnetization curves and the entanglement spectrum and entropy. For
$J_t>0$ a robust ferrimagnetic plateau forms at magnetization per unit
cell $m=1$, whose multiplet entropy reflects how the conserved
magnetization splits between the halves. For $J_t<0$ an $m=0$ plateau
opens and grows with $|J_t|$, while the $m=1$ plateau closes. As
$J_t\to-\infty$ the chain maps onto a spin-$1$ Heisenberg chain with
coupling $J/2$: the $m=0$ width $\Delta h\simeq 0.196$ matches half the
Haldane gap. Exponentially localized spin-$\tfrac{1}{2}$ edge states
and the even-fold degeneracy of the entanglement spectrum confirm the
Haldane character of the $m=0$ phase.
\end{abstract}


\end{frontmatter}

\section{Introduction}
One-dimensional (1D) quantum spin systems provide a paradigmatic
setting to investigate strong correlations and emergent phases in
low dimensions \cite{wen2019a}. In particular, systems with enlarged
unit cells, including trimerized and more general cluster-based
chains \cite{verissimo2025,karlova2018}, have attracted considerable
attention due to the interplay between local cluster formation and
collective quantum behavior.

Trimer chain structures are realized in several quasi-one-dimensional
compounds, including azurite \cite{aimo2009a}, copper-based phosphates
$\mathrm{A}_3\mathrm{Cu}_3(\mathrm{PO}_4)_4$ \cite{belik2005b}, and
$\mathrm{Cu}_3(\mathrm{P}_2\mathrm{O}_6\mathrm{OH})_2$
\cite{hase2020a}. On the theoretical side, these systems have been
extensively investigated within both fermionic and spin frameworks.
In particular, coupled trimer Hubbard chains have been shown to host
rich phase diagrams driven by the interplay between charge and spin
degrees of freedom \cite{macedo1995a,montenegro-filho2005,montenegro-filho2006,rojas2021a,
montenegro-filho2025,saha2025}. Complementary to this, trimerized Heisenberg
models reveal a variety of magnetic phases and excitation spectra,
highlighting the role of frustration and competing exchange
interactions \cite{gu2006b,verkholyak2021a,montenegro-filho2022,
cheng2022,bera2022a,cheng2024a,sen2026,ramos2026}.

Related geometries, such as diamond and branched chains, further enrich
the physical features by incorporating different spin magnitudes and lattice
topologies. These systems include spin-1 \cite{zoshki2026,hashimoto2026},
spin-$\tfrac{1}{2}$ \cite{takano1996a,okamoto2003a,montenegro-filho2008,montenegro-filho2020,li2026},
and mixed spin-$\tfrac{1}{2}$ and spin-1 models \cite{hida2022,
hida2024,hida2026}, as well as branched-chain structures
\cite{karlova2019,souza2020,ghannadan2023}. They exhibit a broad range
of phenomena, including magnetization plateaus, quantum phase
transitions, and exotic excitations such as fractional and composite
quasiparticles. In particular, rich phase diagrams with multiple
phases and critical regions controlled by exchange couplings and
external magnetic fields were observed in a diamond chain
structure~\cite{hida2022,hida2026}.

Within this broader context, a central concept is the Haldane phase of integer-spin
antiferromagnetic chains, characterized by a finite excitation gap and
hidden nonlocal string order. In systems with composite unit cells,
such as trimer chains, effective integer-spin degrees of freedom may
emerge from coupled clusters, giving rise to Haldane-like
phases even when the microscopic spins are half-integer. Such phases
are further characterized by their entanglement
properties, in particular by the degeneracy structure of the
entanglement spectrum, which reflects the underlying
symmetry-protected topological order \cite{wen2019a}.

On the other hand, mixed-spin systems naturally support ferrimagnetic
ground states~\cite{brehmer1997,pati1997a,pati1997b,yamamoto1998,maisinger1998,dasilva2021},
as predicted by the Lieb-Mattis theorem \cite{lieb1962},
which ensures a finite total spin in bipartite systems with unequal
sublattice moments. Among these, the mixed-spin
$1$-$\tfrac{1}{2}$-$\tfrac{1}{2}$ trimer chain provides a natural
setting to investigate the crossover between these two paradigmatic
phases.

In this work, we study a one-dimensional mixed-spin
$1$-$\tfrac{1}{2}$-$\tfrac{1}{2}$ trimer chain using the density matrix
renormalization group (DMRG) method \cite{white1992,
white1993a,schollwock2005,schollwock2007,schollwock2011}. We
investigate the ground-state phases as a function of exchange
couplings and magnetic field, characterizing them through local
magnetization, excitation gaps, and the entanglement spectrum and entropy.
In particular, we show that the entanglement spectrum exhibits the characteristic
degeneracy structure associated with symmetry-protected topological
order \cite{pollmann2010,pollmann2012b}, providing a robust and direct
signature of the Haldane phase, in a well-defined region of the phase diagram.
We further use the entanglement entropy to show that, as
$J_t\to-\infty$, the $m=1$ plateau closes onto the critical point
of the effective spin-$1$ chain at half-saturation, a Tomonaga-Luttinger
liquid of central charge $c=1$.

This work is organized as follows. In Sec.~\ref{sec:model}, we present the model
Hamiltonian and the simulation parameters for the DMRG calculations.
The magnetization curves as a function of magnetic field and local
magnetizations for the $m=1$ plateau are discussed in Sec.~\ref{sec:gs}.
The Haldane phase is characterized in Sec.~\ref{sec:haldane} by the behavior
of the edge states and the degeneracy in the entanglement spectrum.
In Sec.~\ref{sec:ent_m1}, we study the entanglement spectrum and entropy of the $m=1$
plateau, including its closing onto the $c=1$ critical point of the
effective spin-$1$ chain and the entanglement along the ferrimagnetic
multiplet. In Sec.~\ref{sec:summary}, we summarize our main results.

\section{Model and methods}
\label{sec:model}

\begin{figure}[htbp]
    \centering
    \includegraphics[width=0.7\linewidth]{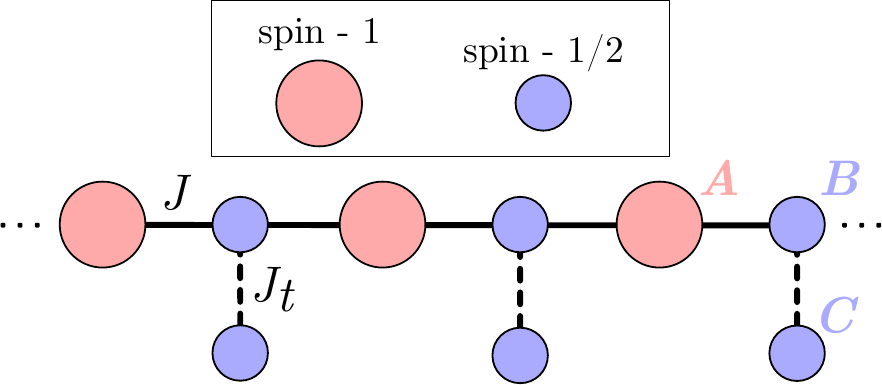}
    \caption{\label{fig:modelo}Schematic representation of the mixed-spin $1$-$\tfrac{1}{2}$-$\tfrac{1}{2}$ trimer chain and the exchange couplings used in the model. Spin-1 degrees of freedom are located on the $A$ sites, whereas the $B$ and $C$ sites carry spin-$\tfrac{1}{2}$. Energy is measured in units of $J\equiv 1$.}
\end{figure}

We consider the mixed-spin trimer chain depicted in Fig.~\ref{fig:modelo}.
Each unit cell $l=1,\ldots,L$ contains a spin-$1$ at site $A$ and two
spin-$\tfrac{1}{2}$ degrees of freedom at sites $B$ and $C$.
The $A$ and $B$ sites form a backbone coupled by the antiferromagnetic
exchange $J>0$, while the $C$ site is attached to $B$ as a side
group through the exchange $J_t$, which we allow to take both positive
and negative values. The Hamiltonian reads
\begin{align}
\mathcal{H} ={}&
J \sum_{l=1}^{L}
\mathbf{S}_{A,l}\cdot \mathbf{s}_{B,l}
+ J \sum_{l=1}^{L-1}
\mathbf{s}_{B,l}\cdot \mathbf{S}_{A,l+1}
\nonumber\\
&+ J_t \sum_{l=1}^{L}
\mathbf{s}_{B,l}\cdot \mathbf{s}_{C,l}
- h \sum_{l=1}^{L}
\left(
S^{z}_{A,l}+s^{z}_{B,l}+s^{z}_{C,l}
\right),
\label{eq:H}
\end{align}
where $\mathbf{S}_{A,l}$ is the spin-$1$ operator at site $A$ of the
$l$th unit cell, $\mathbf{s}_{B,l}$ and $\mathbf{s}_{C,l}$ are the
spin-$\tfrac{1}{2}$ operators at sites $B$ and $C$, and $h$ is a
uniform magnetic field coupled to the total magnetization along the
$z$ axis. Throughout this work, energies are measured in units of the
backbone exchange, $J\equiv 1$.

\begin{figure}[htbp]
    \centering
    \includegraphics[width=0.7\linewidth]{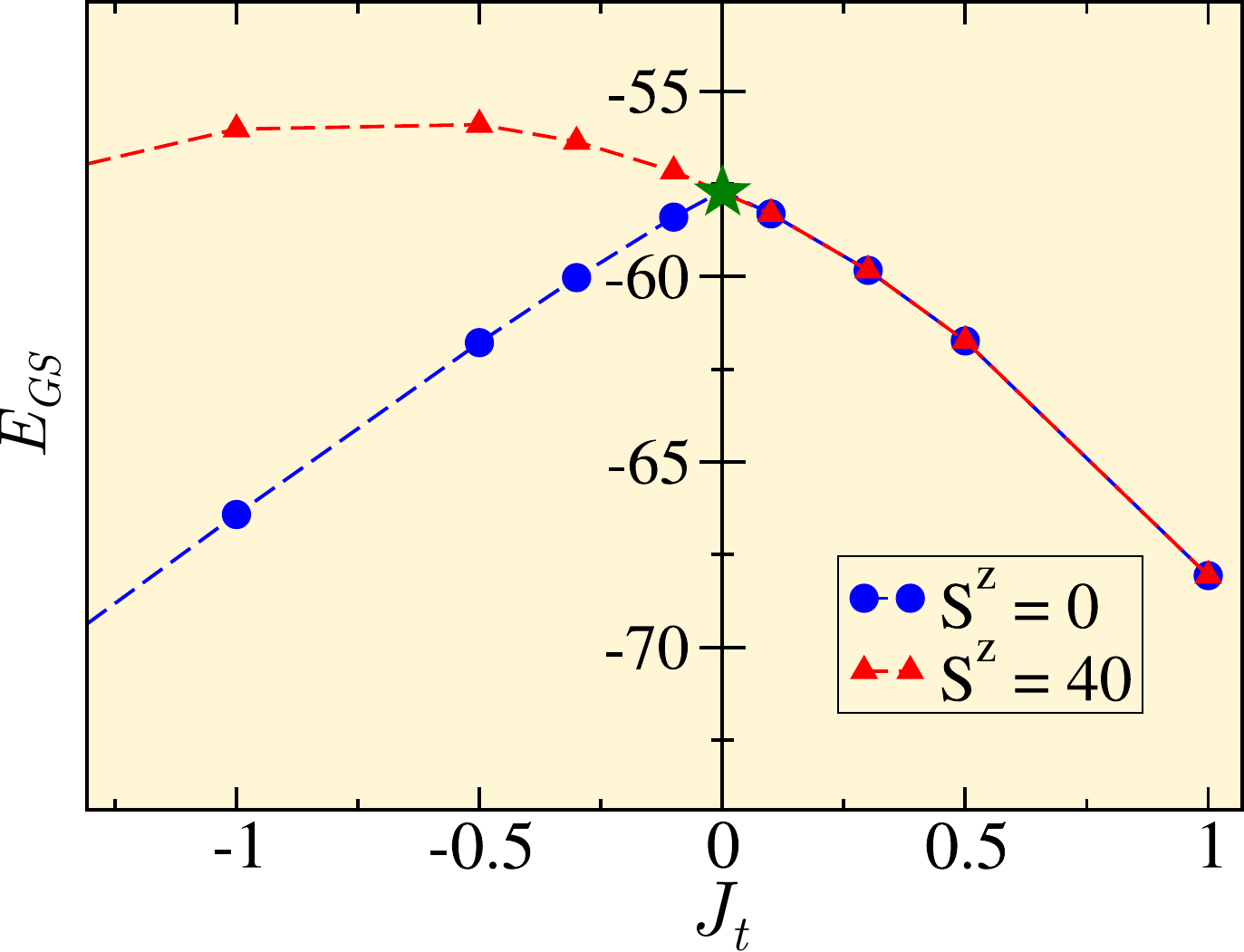}
    \caption{\label{fig:en_jt} Lowest energy as a function of $J_t$ for a trimer chain with $N=120$ sites in the total spin sectors $S^z=0$ and $S^z=L=40$. For $J_t\geq 0$ the two curves coincide, reflecting the SU(2) degeneracy of the ferrimagnetic ground state with total spin $S=L$; for $J_t<0$ they split, with $E(S^z=0)<E(S^z=L)$. At $J_t=0$ the $C$ spins decouple from the backbone, which reduces to an alternating spin-$(1,\tfrac{1}{2})$ chain. The green star marks the ground-state energy of this effective alternating chain with $40$ unit cells, computed independently.}
\end{figure}

We compute the ground state of Eq.~(\ref{eq:H}) using the density
matrix renormalization group (DMRG)
method~\cite{white1992,white1993a,schollwock2005,schollwock2007,schollwock2011}
as implemented in the ITensor library~\cite{itensor}. Calculations
are performed on open chains with $N=3L$ sites, with $L$ ranging from
$40$ up to $200$ unit cells, keeping a maximum bond dimension of
$2000$ and a maximum discarded weight of order $10^{-10}$. The number
of sweeps was varied between $30$ and $70$.

The magnetization curves are obtained from the lowest energy in each
total spin $S^z$ sector at $h=0$, $E(S^z)$, since the Zeeman term in
the Hamiltonian~(\ref{eq:H}) implies $E_h(S^z) = E(S^z) - h S^z$ for
$h\neq 0$. In a finite-size system, the $m(h)$ curve is composed of
finite-size steps of width $\Delta h(S^z)$ at total spin $S^z$.
Defining $h_{S^z+}$ and $h_{S^z-}$ as the extreme points of these
steps, such that $\Delta h(S^z)=h_{S^z+}-h_{S^z-}$, we have
\begin{equation}
h_{S^z\pm} = \pm\left[E(S^z\pm 1)-E(S^z)\right].
\label{eq:hpm}
\end{equation}
If $S^z$ does not correspond to
a thermodynamic-limit magnetization plateau, then $\Delta
h(S^z)\rightarrow 0$ as $L\rightarrow\infty$; otherwise, $\Delta
h(S^z)\neq 0$ in the same limit.

\section{Ground state energy and magnetization curves}
\label{sec:gs}

We begin the discussion of the ground-state phases by examining the
dependence of the lowest energy in each total-$S^z$ sector on the
side-coupling $J_t$. In Fig.~\ref{fig:en_jt} we plot $E(S^z)$ as a
function of $J_t$ for a chain with $N=120$ sites in the $S^z=0$ and
$S^z=L=40$ sectors. For $J_t>0$ the ground state is the
ferrimagnetic state with total spin $S=L=40$ and magnetization $m=1$
per unit cell; by SU(2) symmetry, the ground-state energy is
degenerate across all sectors with $|S^z|\leq L$, and
in particular the $E(S^z=0)$ and $E(S^z=L)$ curves coincide in this
regime. As $J_t$ is reduced and crosses zero, the two curves remain coincident
down to $J_t=0$, where the $C$ spins decouple from the backbone and
the remaining $A$--$B$ chain reduces to an alternating
spin-$(1,\tfrac{1}{2})$ Heisenberg chain. The green star in
Fig.~\ref{fig:en_jt} marks the ground-state energy of this effective
alternating chain with $40$ unit cells, computed independently, and
matches the common value of $E(S^z=0)$ and $E(S^z=L)$ at $J_t=0$
within the numerical precision. For $J_t<0$ the two curves split,
with $E(S^z=0)<E(S^z=L)$, so that the $m=1$ ferrimagnetic state is no
longer the lowest-energy state; the actual ground state is a singlet,
as discussed below.

We define the magnetization per unit cell as
\begin{equation}
m \equiv \frac{1}{L}\sum_{l=1}^{L}\left\langle S^{z}_{A,l}+s^{z}_{B,l}+s^{z}_{C,l}\right\rangle,
\label{eq:m}
\end{equation}
where $\langle\cdots\rangle$ denotes the ground-state expectation value.

\begin{figure}[htbp]
    \centering
    \includegraphics[width=0.7\linewidth]{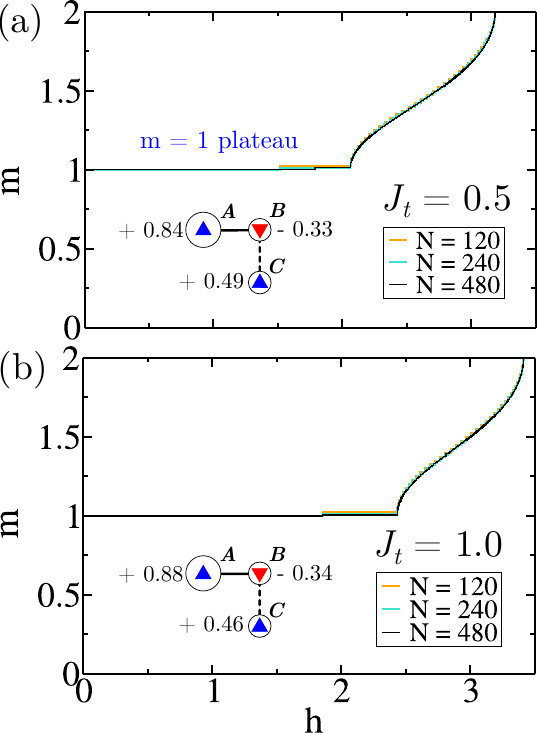}
    \caption{\label{fig:mag_jtpos}Magnetization per unit cell, $m$, as a function of the magnetic field $h$ for several system sizes $N$ and for (a) $J_t=0.5$ and (b) $J_t=1.0$. An $m=1$ ferrimagnetic plateau is observed for $J_t>0$. Insets show the site-resolved ground-state magnetizations $\langle S^z_{A}\rangle$, $\langle s^z_{B}\rangle$, and $\langle s^z_{C}\rangle$ within the plateau, measured at the center of the open chain for $N=480$, evidencing ferrimagnetic alignment between the $(A+C)$ and $B$ sublattices.}
\end{figure}

In Fig.~\ref{fig:mag_jtpos} we present the magnetization curves
$m(h)$ for $J_t>0$, with $J_t=0.5$ in panel (a) and $J_t=1.0$ in
panel (b). For both couplings the curves display a robust plateau at
$m=1$, whose width is independent of system size within our
resolution, confirming its persistence in the thermodynamic limit. The
lattice is bipartite, with $A$ and $C$ sites on one sublattice and
$B$ sites on the other, since $A$ and $C$ couple only to $B$ and the
bonds $A$--$A$, $B$--$B$, and $A$--$C$ are absent. For $J_t>0$ all
exchange couplings are antiferromagnetic, so that the conditions of
the Lieb-Mattis theorem~\cite{lieb1962} are met, and the theorem
fixes the ground-state spin per unit cell at $|S_A+s_C-s_B|=1$, in
agreement with the observed plateau. The site-resolved magnetizations
within the plateau, shown in the insets, confirm this picture:
$\langle S^z_A\rangle$ and $\langle s^z_C\rangle$ are positive, while
$\langle s^z_B\rangle$ is negative, evidencing the ferrimagnetic
alignment between the $(A+C)$ and $B$ sublattices.

The Oshikawa--Yamanaka--Affleck condition for the appearance of
magnetization plateaus in spin chains~\cite{oshikawa1997},
$S_{\mathrm{uc}}-m\in\mathbb{Z}$ with $S_{\mathrm{uc}}=S_A+s_B+s_C=2$ the
total spin per unit cell, allows plateaus at $m=0$, $m=1$, and at the
saturation value $m=2$. The $m=1$ plateau is the one realized for
$J_t>0$; as shown below, an $m=0$ plateau also develops for $J_t<0$.
All occur without spontaneous breaking of translational symmetry.

\begin{figure}[htbp]
    \centering
    \includegraphics[width=0.45\linewidth]{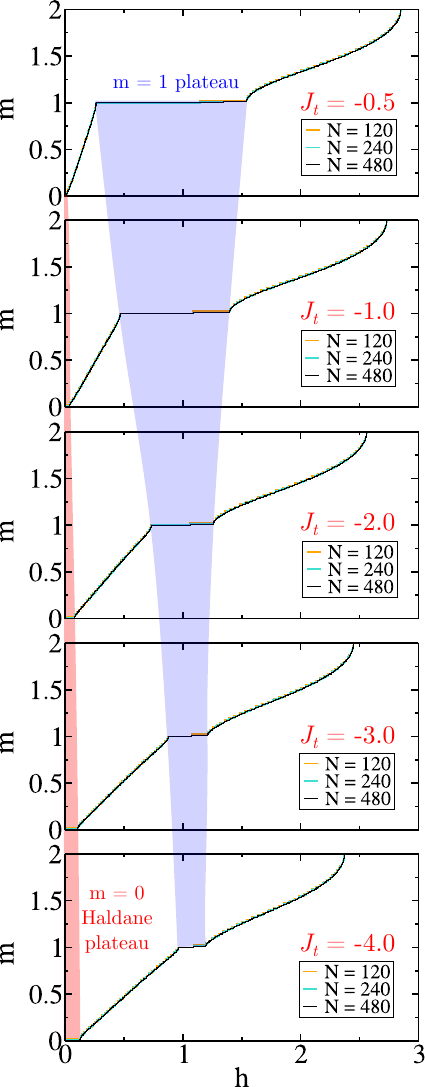}
    \caption{\label{fig:mag_jtneg}Magnetization per unit cell, $m$, as a function of the magnetic field $h$ for several system sizes $N$ and for $J_t<0$: $J_t=-0.5$, $-1.0$, $-2.0$, $-3.0$, and $-4.0$ (top to bottom). The shaded regions highlight the evolution of the $m=1$ plateau as $J_t$ decreases, and the emergence and growth of an $m=0$ plateau (Haldane gap) for $J_t<0$.}
\end{figure}

For $J_t<0$ the $B$--$C$ exchange becomes ferromagnetic and the
conditions of applicability of the Lieb-Mattis theorem are no
longer met;
the nature of the ground state in this regime must therefore be
established directly from the numerical data.
Figure~\ref{fig:mag_jtneg} shows the magnetization curves for $J_t=-0.5$,
$-1.0$, $-2.0$, $-3.0$, and $-4.0$. As $J_t$ becomes more negative,
the $m=1$ plateau shrinks while a new plateau at $m=0$ opens up and
grows monotonically, signaling the development of a Haldane-type
gapped phase whose nature is investigated in detail in
Sec.~\ref{sec:haldane}.

\begin{figure}[htbp]
    \centering
    \includegraphics[width=0.7\linewidth]{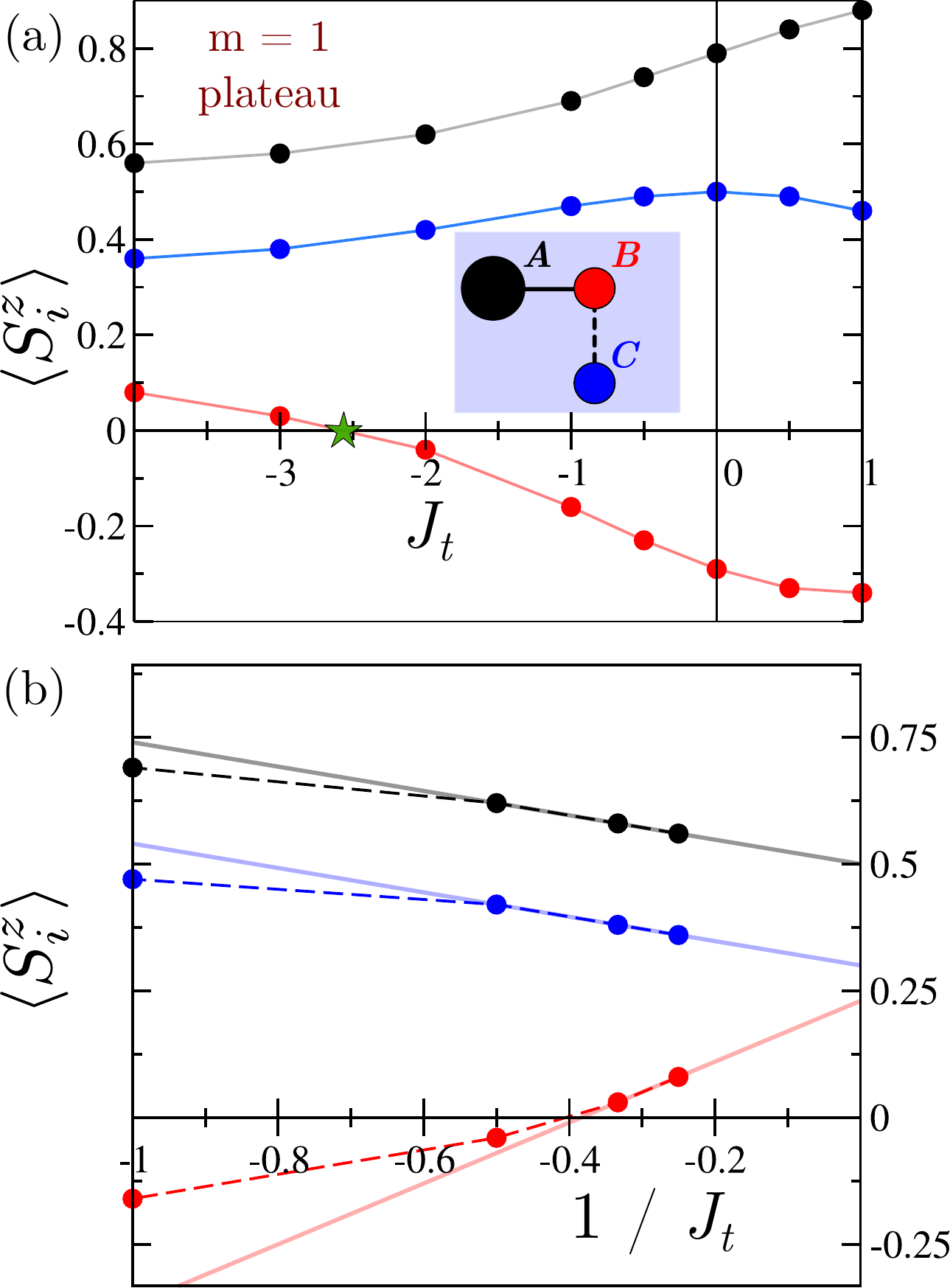}
    \caption{\label{fig:localsz_jt}Site-resolved ground-state magnetizations $\langle S^z_A\rangle$, $\langle s^z_B\rangle$, and $\langle s^z_C\rangle$ within the $m=1$ plateau, measured at the center of an open chain with $N=480$. (a) Local magnetizations as functions of $J_t$; the star marks the coupling where the $B$-sublattice magnetization vanishes, $\langle s^z_B\rangle=0$. (b) Same data plotted versus $1/J_t$ to probe the $J_t\to -\infty$ limit. The solid lines are linear fits to the two largest-$|J_t|$ points, yielding extrapolated values $\langle S^z_A\rangle= 0.50$, $\langle s^z_B\rangle\simeq 0.23$, and $\langle s^z_C\rangle\simeq 0.30$ at $1/J_t=0$.}
\end{figure}

The reorganization of the local magnetizations along the
negative-$J_t$ axis is shown in Fig.~\ref{fig:localsz_jt}. Panel (a)
displays $\langle S^z_A\rangle$, $\langle s^z_B\rangle$, and
$\langle s^z_C\rangle$ within the $m=1$ plateau as functions of
$J_t\leq 0$. As $|J_t|$ grows, the $B$ sublattice magnetization changes sign at
the value marked by the star, where $\langle s^z_B\rangle=0$, and
for sufficiently negative $J_t$ the three sublattices $A$, $B$, and
$C$ are all polarized in the same direction.
This inversion of the local pattern, although it leaves the
integrated $m=1$ plateau intact, anticipates the qualitative
reorganization of the ground state at lower fields, where the $m=0$
Haldane plateau sets in. Panel (b) displays the same data versus
$1/J_t$ to access the $J_t\to-\infty$ limit; the solid lines are
linear fits to the two largest-$|J_t|$ points, yielding the
extrapolated values $\langle S^z_A\rangle=0.50$,
$\langle s^z_B\rangle\simeq 0.23$, and $\langle s^z_C\rangle\simeq
0.30$ at $1/J_t=0$. We note that in this limit the $B$--$C$ pair is
locked into a triplet by the strong ferromagnetic $J_t$ and behaves
as an effective spin-$1$, while the spin-$1$ at site $A$ is
antiferromagnetically coupled to it through $J$; the system thus
maps onto a uniform spin-$1$ Heisenberg chain, in which $m=1$ in
the original trimer chain corresponds to half-saturation. Although
the individual extrapolations of $\langle s^z_B\rangle$ and
$\langle s^z_C\rangle$ yield distinct values, $\simeq 0.23$ and
$\simeq 0.30$, respectively, reflecting the different roles of $B$
(directly coupled to the backbone) and $C$ (coupled only through
$B$) at finite $1/|J_t|$, they add up to $\simeq 0.53$, close to the
half-saturation value $\langle S^z_A\rangle\simeq 0.50$ of the
effective spin-$1$ site and confirming the emergence of an effective
spin-$1$ chain in this limit. The small excess of this sum over
$0.50$, like the difference between the individual $B$ and $C$
extrapolations, is most likely a finite-size effect that should be
reduced for larger chains.

\subsection{Plateau widths behavior for $J_t<0$}

We now examine quantitatively how the magnetic plateaus evolve along
the negative-$J_t$ axis, starting with the $m=1$ plateau and then
turning to the $m=0$ Haldane plateau. To characterize the $m=1$
plateau we use the lower and upper finite-size edges of the
magnetization step at $S^z=L$, Eq.~(\ref{eq:hpm}). The lower edge is
\begin{equation}
h_{c1} \equiv E(L)-E(L-1).
\label{eq:hc1}
\end{equation}
The upper edge in the open chain is affected by a localized boundary
mode of the kind commonly found in ferrimagnetic spin
chains~\cite{montenegro-filho2020,dasilva2021}, and we therefore
distinguish the finite-size step,
\begin{equation}
h^{(\mathrm{edge})} \equiv E(L+1)-E(L),
\label{eq:hedge}
\end{equation}
from the bulk gap obtained by skipping the boundary contribution,
\begin{equation}
h_{c2} \equiv E(L+2)-E(L+1).
\label{eq:hc2}
\end{equation}
The thermodynamic plateau width is $\Delta h(m=1)=h_{c2}-h_{c1}$,
evaluated from the $N\to\infty$ extrapolation of the finite-size critical
fields.

\begin{figure}[htbp]
    \centering
    \includegraphics[width=0.85\linewidth]{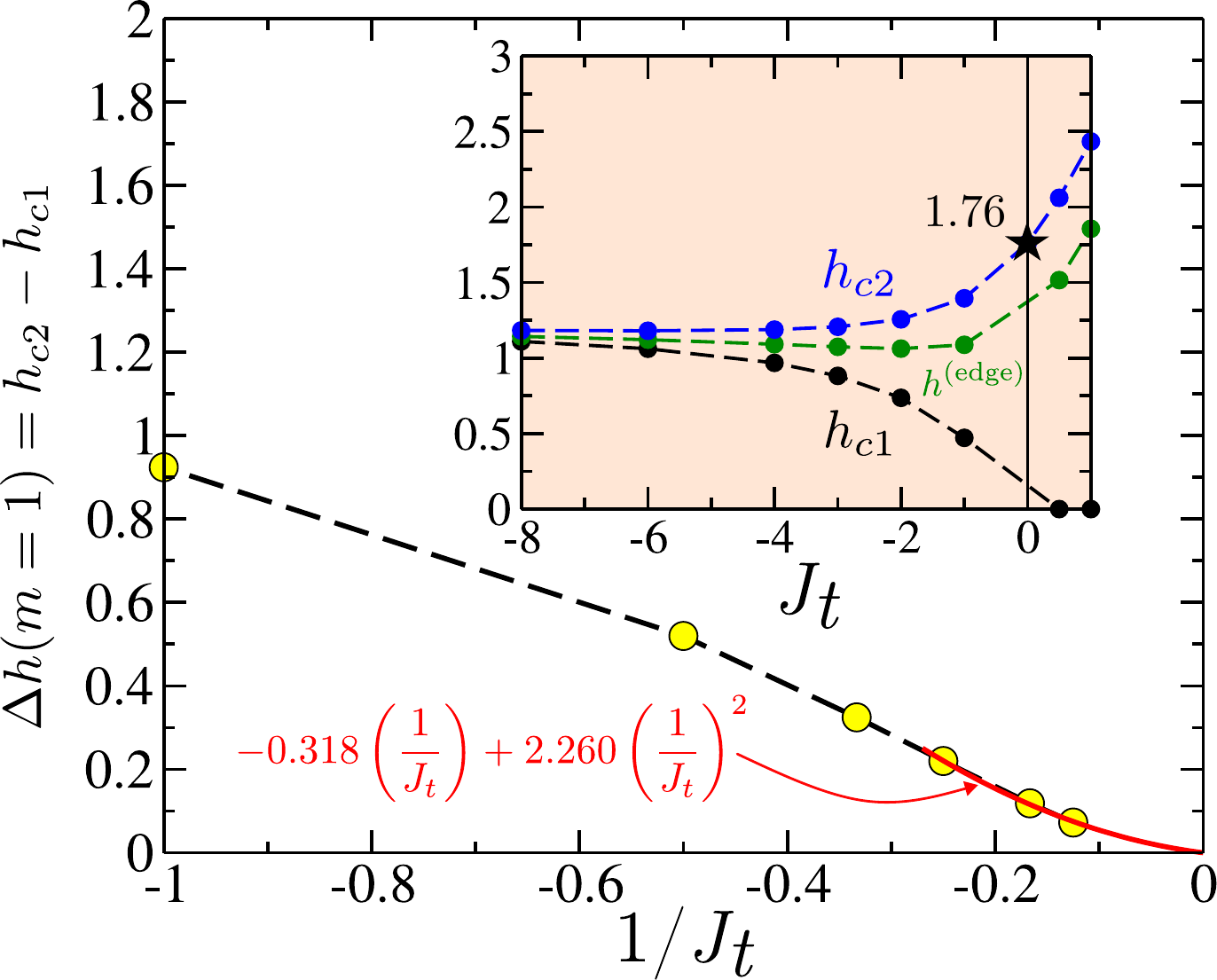}
    \caption{\label{fig:hm1_width}Main panel: width of the $m=1$
    plateau, $\Delta h(m=1)=h_{c2}-h_{c1}$, as a function of $1/J_t$
    for $J_t<0$, extrapolated to $N\to\infty$ linearly in $1/L$ from
    system sizes $N=180$, $300$, and $420$. The dashed line is a
    guide to the eye; the red solid curve is a two-parameter fit
    $\Delta h(m=1)=-0.318\,(1/J_t)+2.260\,(1/J_t)^2$ to the three
    largest-$|J_t|$ data points, with no constant term, so that
    $\Delta h(m=1)\to 0$ as $1/J_t\to 0$ ($J_t\to-\infty$). Inset:
    critical fields $h_{c1}$, $h^{(\mathrm{edge})}$, and $h_{c2}$
    (see text) versus $J_t$, including the $J_t>0$ side, also
    extrapolated to $N\to\infty$ from the same set of sizes. The
    star at $h\simeq 1.76$, $J_t=0$, marks the right-edge gap of the
    $m=\tfrac{1}{2}$ ferrimagnetic plateau of the alternating
    spin-$(1,\tfrac{1}{2})$
    chain~\cite{brehmer1997,pati1997a,pati1997b,yamamoto1998,maisinger1998}.}
\end{figure}

In Fig.~\ref{fig:hm1_width} we plot $\Delta h(m=1)$ as a function of
$1/J_t$, together with $h_{c1}$, $h^{(\mathrm{edge})}$, and $h_{c2}$
versus $J_t$ in the inset. The width decreases monotonically as
$|J_t|$ grows. At $J_t=0$ the $C$ spins decouple from the backbone
and the $m=1$ state of the trimer chain consists of the
ferrimagnetic ground state of the alternating
spin-$(1,\tfrac{1}{2})$ chain, with $m_{\mathrm{alt}}=\tfrac{1}{2}$,
plus the fully polarized $C$ spin contributing the remaining
$\tfrac{1}{2}$. Accordingly, $h_{c1}\to 0$ as $J_t\to 0^-$, since
flipping the decoupled $C$ spin is free, while $h_{c2}$ approaches
the right-edge gap of the $m=\tfrac{1}{2}$ plateau of the alternating
spin-$(1,\tfrac{1}{2})$
chain~\cite{brehmer1997,pati1997a,pati1997b,yamamoto1998,maisinger1998},
marked by the star at $h\simeq 1.76$ in the inset. Across the
negative-$J_t$ axis the edge field stays below the bulk gap,
$h^{(\mathrm{edge})}<h_{c2}$ (inset), reflecting the open-boundary
mode introduced above; the two fields merge as $J_t\to-\infty$,
where the trimer chain maps onto a uniform spin-$1$ Heisenberg
chain (see below). The latter has no plateau at half-saturation,
so that $\Delta h(m=1)\to 0$ as $1/J_t\to 0$. The red solid curve in
Fig.~\ref{fig:hm1_width} is a two-parameter fit
$\Delta h(m=1)=a_0(1/J_t)+a_1(1/J_t)^2$, with no constant term, to
the three largest-$|J_t|$ data points, making the asymptotic closure
of the plateau manifest.

\begin{figure}[htbp]
    \centering
    \includegraphics[width=0.8\linewidth]{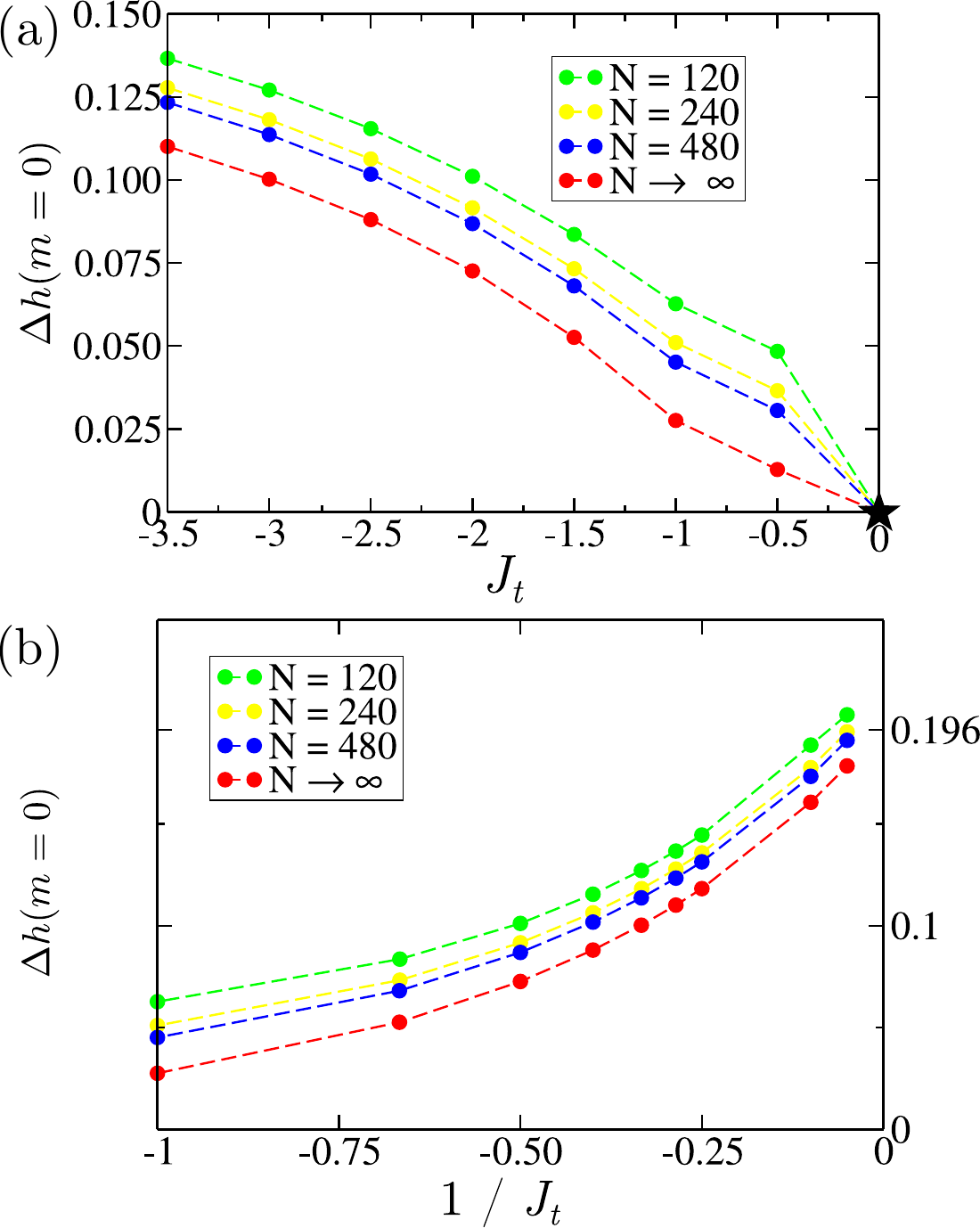}
    \caption{\label{fig:hm0_width}Width of the $m=0$ Haldane plateau,
    $\Delta h(m=0)$, for $J_t<0$, computed for system sizes $N=120$,
    $240$, and $480$, together with the $N\to\infty$ extrapolation
    linear in $1/L$. (a) $\Delta h(m=0)$ versus $J_t$: the plateau is
    absent at $J_t=0$ (star) and opens continuously as $J_t$ becomes
    negative. (b) Same data versus $1/J_t$: the $N\to\infty$
    extrapolation reaches $\Delta h(m=0)\simeq 0.196$ at $1/J_t=0$,
    matching half of the Haldane gap of the spin-$1$ Heisenberg
    chain~\cite{white1993b}.}
\end{figure}

The $m=0$ Haldane plateau, shown in Fig.~\ref{fig:hm0_width}, is
absent at $J_t=0$ [star in panel (a)], opens continuously as $J_t$
becomes negative, and grows monotonically with $|J_t|$. The linear
extrapolation in $1/L$ shown in panel (b) yields
$\Delta h(m=0)\simeq 0.196$ at $1/J_t=0$, in good agreement with the
prediction of the effective spin-$1$ Heisenberg chain that emerges
in the $|J_t|\to\infty$ limit.

To make this correspondence explicit,
we project the $B$--$C$ pair onto its exact triplet, identifying
\begin{align}
\mathbf{s}_{B,l} &= \mathbf{T}_{x=2l}/2, \nonumber\\
\mathbf{s}_{C,l} &= \mathbf{T}_{x=2l}/2, \\
\mathbf{S}_{A,l} &\to \mathbf{T}_{x=2l-1}, \nonumber
\end{align}
where $\mathbf{T}_{x}$ is a spin-$1$ operator and
$x=1,2,\ldots,2L$ enumerates effective sites along the chain
[the resulting geometry coincides with the labeling of
Fig.~\ref{fig:edge_profile}(a)]. Substituting these identifications
into Eq.~(\ref{eq:H}), the side-coupling term reduces to a constant
in the triplet subspace, $\mathbf{s}_{B,l}\cdot\mathbf{s}_{C,l}=\tfrac{1}{4}$,
which is absorbed into an overall energy shift. The original
Hamiltonian then reduces to
\begin{equation}
\mathcal{H}_{\mathrm{eff}} = \frac{J}{2}\sum_{x=1}^{2L-1}
\mathbf{T}_{x}\cdot\mathbf{T}_{x+1} + \text{const.},
\label{eq:Heff}
\end{equation}
that is, a uniform spin-$1$ Heisenberg chain with effective coupling
$J/2$. The $m=0$ plateau width in this limit is therefore expected to
match the Haldane gap of the spin-$1$ Heisenberg chain rescaled by
$1/2$, $\Delta_{\mathrm{Haldane}}/2\simeq 0.41/2\simeq 0.2$~\cite{white1993b},
in excellent agreement with the extrapolated value of
Fig.~\ref{fig:hm0_width}(b).

The same effective spin-$1$ chain also fixes the field at which the
$m=1$ plateau of the trimer closes as $J_t\to-\infty$. In the
effective chain, $m=1$ per trimer unit cell corresponds to
$m_{\mathrm{eff}}=\tfrac{1}{2}$ per effective spin-$1$ site, that is,
half-saturation, which is reached at the well-established field
$H_{S=1}(m_{\mathrm{eff}}=\tfrac{1}{2})\simeq 2.5$ in units of the
spin-$1$ chain coupling~\cite{sakai1991}. With effective coupling
$J/2$, this translates into
$h(m=1,J_t\to-\infty)\simeq (J/2)\times 2.5\simeq 1.25$. The inset
of Fig.~\ref{fig:hm1_width} is consistent with this estimate: as
$1/J_t\to 0$, the three fields $h_{c1}$, $h^{(\mathrm{edge})}$, and
$h_{c2}$ converge towards $h\simeq 1.2\text{--}1.25$, jointly
encoding the closure of the plateau, $\Delta h(m=1)\to 0$, and the
disappearance of the boundary mode,
$h^{(\mathrm{edge})}\to h_{c2}$, at the half-saturation field of
the underlying spin-$1$ Heisenberg chain.

\section{Haldane phase}
\label{sec:haldane}

We now characterize the gapped $m=0$ phase that develops for
$J_t<0$. The discussion of Sec.~\ref{sec:gs} established that, in
the $J_t\to-\infty$ limit, the $B$--$C$ pair locks into a triplet
and the trimer chain reduces to a uniform spin-$1$ Heisenberg chain,
whose ground state is the prototypical Haldane state. Here we show
that this Haldane character extends to the whole $m=0$ region of the
phase diagram explored in this work, by examining two complementary
signatures: the spatial profile of the local magnetization in open
chains, which exposes fractional spin-$\tfrac{1}{2}$ edge states, and
the degeneracy structure of the entanglement spectrum, which provides
a direct fingerprint of the underlying symmetry-protected
topological order~\cite{pollmann2010,pollmann2012b}.

It is instructive to view the $m=0$ ground state through an
AKLT-like picture~\cite{aklt1987,aklt1988} adapted to the trimer
geometry. Decomposing the spin-$1$ at site $A$ into two virtual
spin-$\tfrac{1}{2}$ degrees of freedom, each virtual $\tfrac{1}{2}$
tends to lock into a singlet with one of the real spin-$\tfrac{1}{2}$
operators of the same unit cell, $\mathbf{s}_{B,l}$ or
$\mathbf{s}_{C,l}$, in close analogy with the valence-bond-solid
structure of the spin-$1$ chain, except that the singlet partners
here are real lattice spins rather than virtual halves of
neighboring spin-$1$ sites. In an open chain, this construction
naturally leaves uncompensated spin-$\tfrac{1}{2}$ degrees of freedom
at the two boundaries, which manifest themselves as the
fractional edge states discussed below.

\begin{figure}[htbp]
    \centering
    \includegraphics[width=\linewidth]{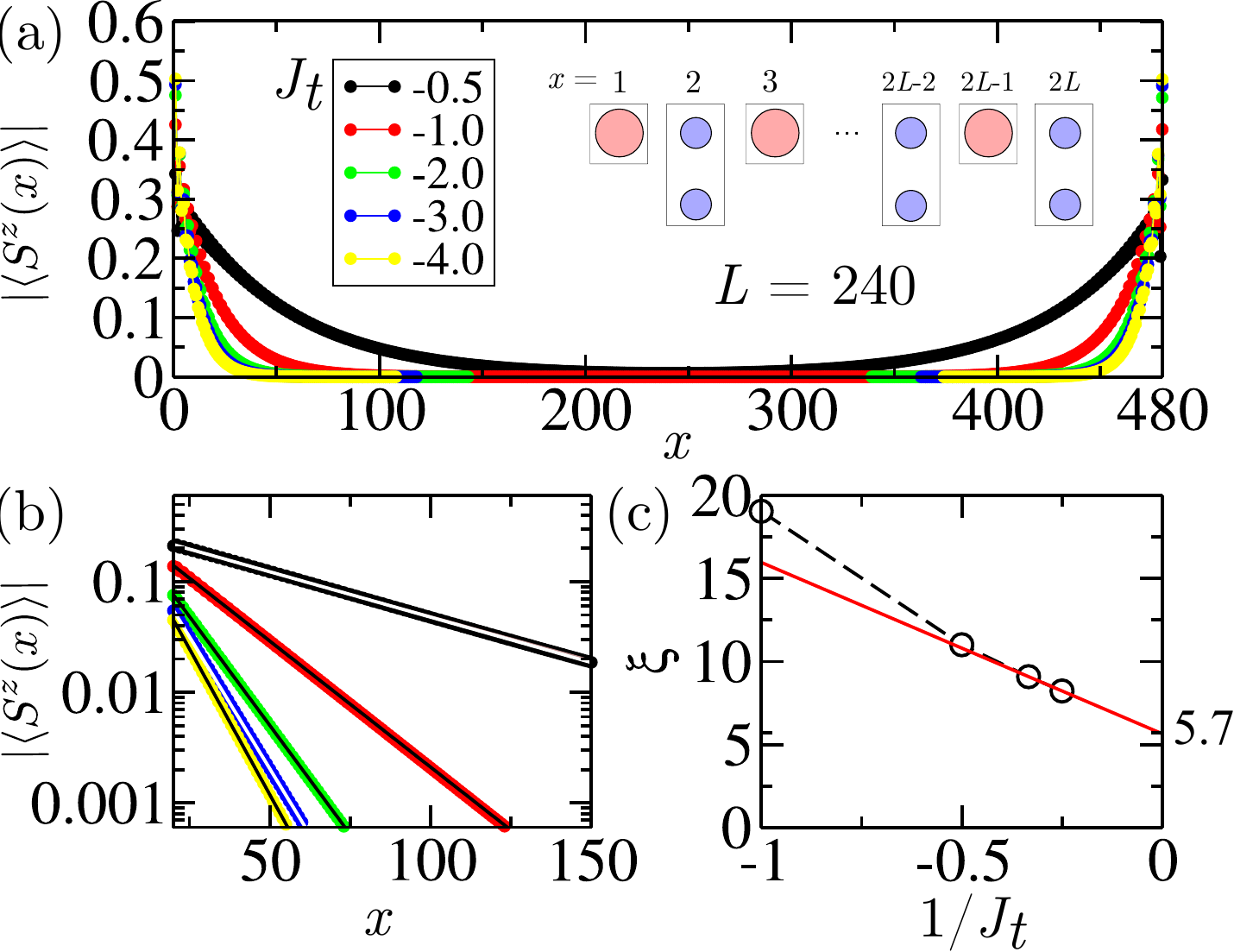}
    \caption{\label{fig:edge_profile}Spatial profile of the local magnetization magnitude in the $m=0$ (Haldane) phase for several values of $J_t<0$, computed in the $S^z=0$ sector. (a) $|\langle S^z(x)\rangle|$ as a function of position $x$ for an open chain (illustrated in the inset): odd $x$ correspond to spin-1 sites, while even $x$ denote the sum of the two spin-$\tfrac{1}{2}$ degrees of freedom within the same unit cell. (b) Semi-log plot of $|\langle S^z(x)\rangle|$ near the left edge; solid lines are exponential fits $|\langle S^z(x)\rangle|\sim \exp(-x/\xi)$, defining the decay length $\xi$. (c) Decay length $\xi$ as a function of $1/J_t$; linear extrapolation to $1/J_t=0$ yields $\xi(J_t\to -\infty)\simeq 5.7$.}
\end{figure}

In Fig.~\ref{fig:edge_profile}(a) we display the magnitude of the
local magnetization, $|\langle S^z(x)\rangle|$, as a function of the
position $x$ along the open chain in the $S^z=0$ sector,
for several values of $J_t<0$. We adopt a coarse-grained
labeling in which odd $x$ corresponds to the spin-$1$ at site $A$
and even $x$ to the sum
$\langle s^z_{B,l}\rangle+\langle s^z_{C,l}\rangle$ of the two
spin-$\tfrac{1}{2}$ within the same unit cell, so that $x$ effectively
indexes the sites of the underlying spin-$1$-like chain. The profiles
$|\langle S^z(x)\rangle|$ are localized at the boundaries and
decay exponentially towards the bulk, exhibiting the characteristic
two-sublattice alternation of an AKLT-like edge mode. The
integrated boundary magnetization carries opposite signs at the
two ends and approaches $\pm\tfrac{1}{2}$ as $|J_t|$ grows,
consistent with two effective spin-$\tfrac{1}{2}$ edge states that
become essentially decoupled in this limit. For $J_t$ closer to
zero, the larger bulk correlation length enhances the residual
coupling between the edge modes through the bulk, reducing
$|\langle S^z\rangle|$ at the boundaries; the saturation observed
at large $|J_t|$ therefore reflects the decoupling of the two
edge spin-$\tfrac{1}{2}$ as $\xi$ shrinks [see
Fig.~\ref{fig:edge_profile}(c)].

The semi-log plot of Fig.~\ref{fig:edge_profile}(b) makes the
exponential character of the boundary localization explicit: the
data near the left edge follow the form
$|\langle S^z(x)\rangle|\sim e^{-x/\xi}$, defining a decay length
$\xi$. The values of $\xi$ extracted from the exponential fits are
shown in Fig.~\ref{fig:edge_profile}(c) as a function of $1/J_t$.
A linear extrapolation of the two largest-$|J_t|$ points yields
$\xi(J_t\to-\infty)\simeq 5.7$. This edge-state localization length
coincides with the bulk correlation length of the spin-$1$ Heisenberg
chain, $\xi\simeq 6$~\cite{white1993b}, as expected in the Haldane
phase, where the localization of the boundary mode and the decay of
bulk correlations are both controlled by the same Haldane gap. The
agreement thus provides a quantitative check of the mapping between the
strongly-ferromagnetic-side-coupling limit and the Haldane chain.

\begin{figure}[htbp]
    \centering
    \includegraphics[width=\linewidth]{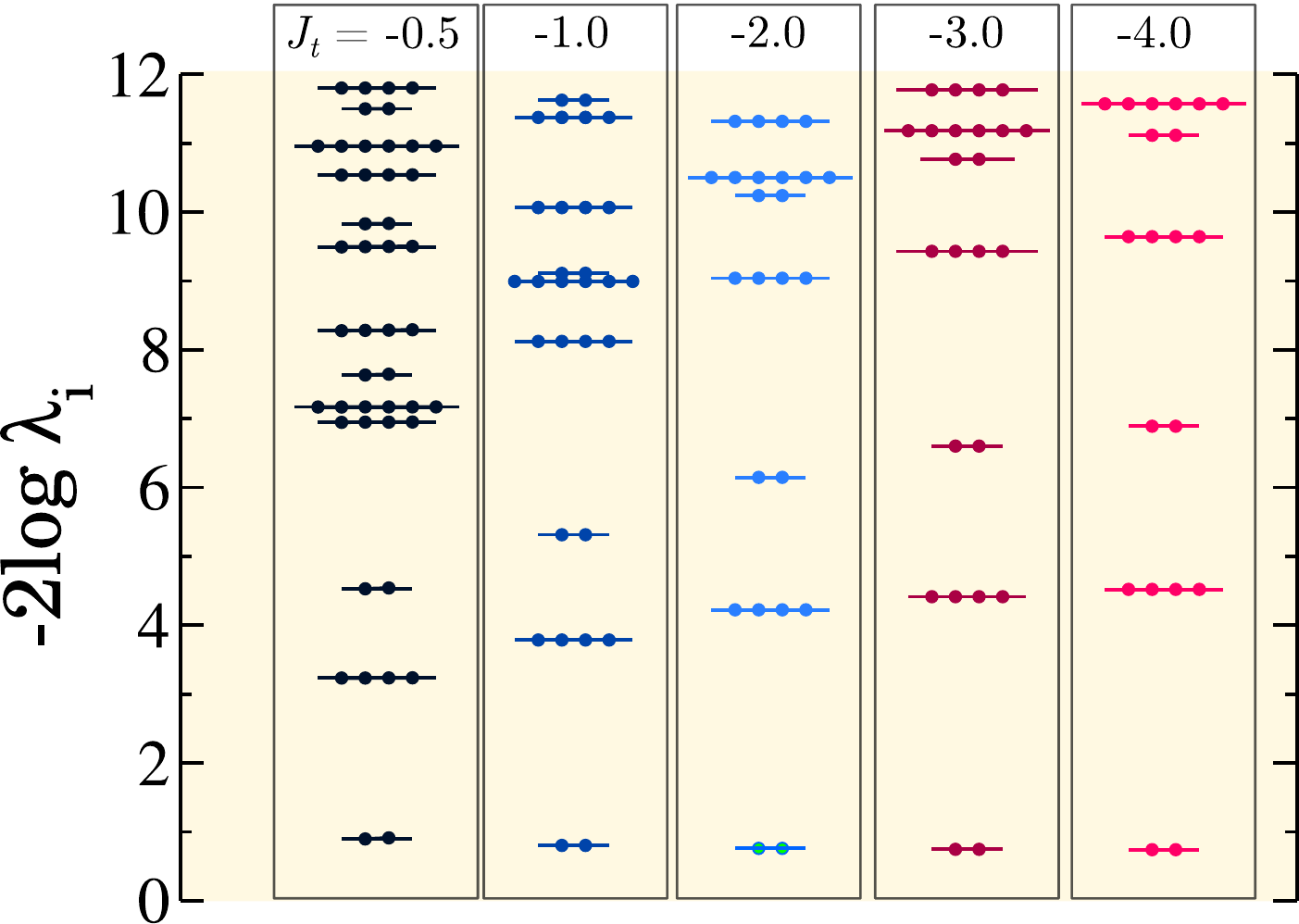}
    \caption{\label{fig:ent_spec}Entanglement spectrum in the $m=0$ (Haldane) phase for the values of $J_t$ indicated in the figure. The levels correspond to $-2\ln(\lambda_i)$, where $\lambda_i$ are the Schmidt coefficients of the ground state for a bipartition at the center of the open chain.}
\end{figure}

A direct diagnostic of the topological character of this phase is
provided by the entanglement spectrum. Following
Refs.~\cite{pollmann2010,pollmann2012b}, we compute the Schmidt
decomposition of the ground state across a bipartition at the center
of the open chain and report the entanglement levels
$\epsilon_i\equiv -2\ln\lambda_i$, where $\lambda_i$ are the Schmidt
coefficients. The result is shown in Fig.~\ref{fig:ent_spec} for
several values of $J_t<0$. For all couplings examined, the entanglement
levels organize into pairs of essentially degenerate levels, the
trademark even-fold degeneracy associated with symmetry-protected
topological order. This degeneracy persists across
the whole range of $J_t$ explored, confirming that the $m=0$ phase
identified from the magnetization curves of
Fig.~\ref{fig:mag_jtneg} is adiabatically connected to the
Haldane phase of the spin-$1$ chain attained in the $J_t\to-\infty$
limit.

\section{Entanglement in the $m=1$ plateau}
\label{sec:ent_m1}

\begin{figure}[htbp]
    \centering
    \includegraphics[width=0.9\linewidth]{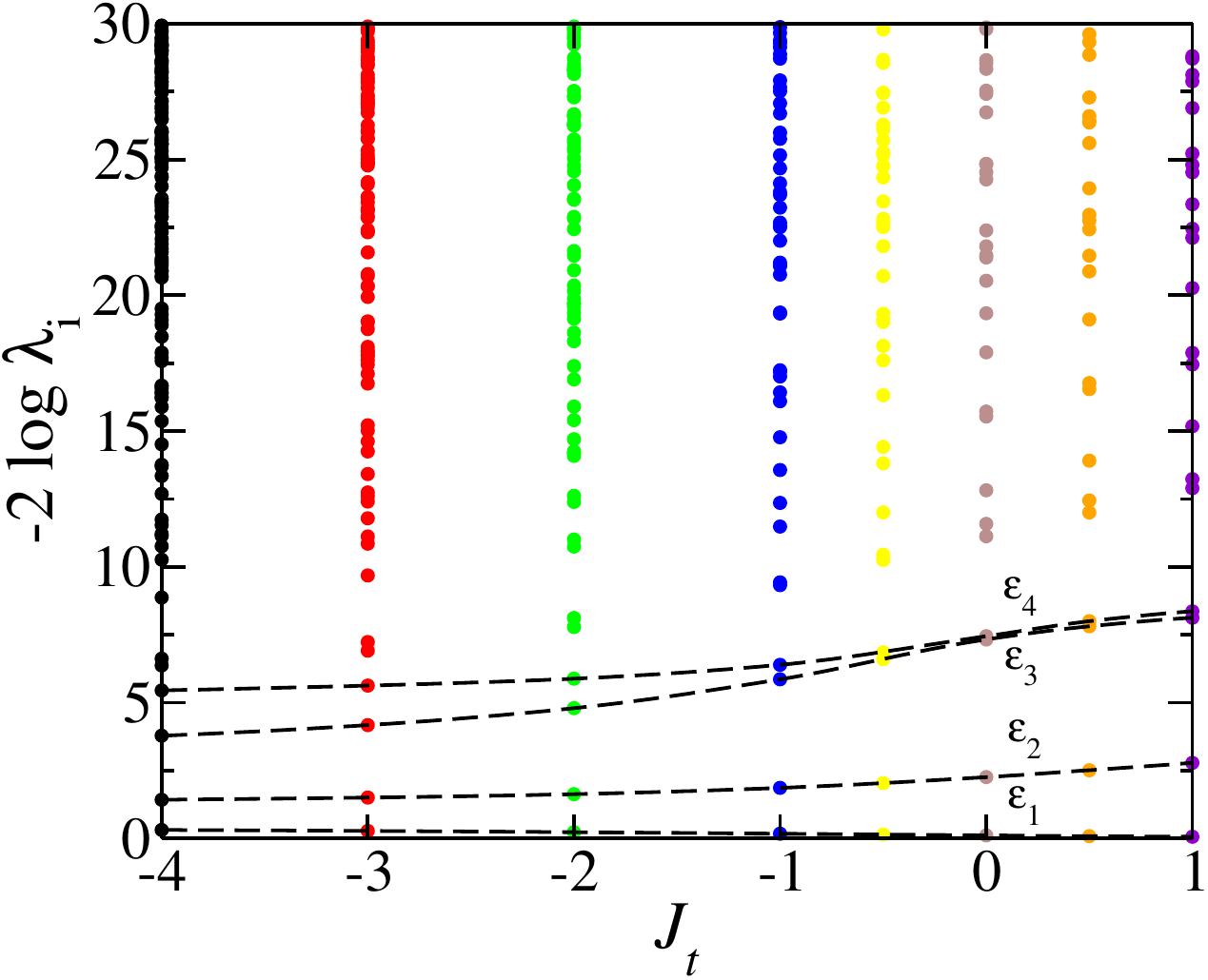}
    \caption{\label{fig:ent_spec_m1}Entanglement spectrum of the $m=1$
    plateau ($S^z=L$ sector) of a chain with $L=200$ unit cells, for
    the values of $J_t$ indicated, with
    levels $-2\ln(\lambda_i)$ obtained from the Schmidt coefficients
    $\lambda_i$ of the ground state for a bipartition at the center of
    the open chain. The dashed lines are guides to the eye tracking the
    four lowest levels $\epsilon_1$--$\epsilon_4$ across $J_t$. Unlike
    the $m=0$ Haldane phase (Fig.~\ref{fig:ent_spec}), these lowest
    levels are non-degenerate.}
\end{figure}

It is instructive to begin with the entanglement spectrum of the
$m=1$ plateau, computed exactly as in the $m=0$ phase
(Sec.~\ref{sec:haldane}) but now in the $S^z=L$ sector: we take the
Schmidt decomposition of the ground state across the central bond of
the open chain and report the entanglement levels
$\epsilon_i\equiv-2\ln\lambda_i$. Figure~\ref{fig:ent_spec_m1} shows
these levels for several values of $J_t$, from the
ferrimagnetic-side-coupling regime ($J_t>0$) down to $J_t=-4$. In
sharp contrast to the $m=0$ Haldane phase (Fig.~\ref{fig:ent_spec}),
the lowest entanglement levels are now \emph{non-degenerate}: the
even-fold degeneracy that is the hallmark of symmetry-protected
topological order is absent throughout the range of $J_t$ examined,
the extrapolated lowest levels forming a single ladder
($\epsilon_{1\ldots4}\to0.48,\,1.17,\,2.43,\,4.56$). The $m=1$ plateau
therefore carries no Haldane-type topological order, as expected since
the finite magnetization explicitly breaks the spin-rotation symmetry
that protects it. As $J_t$ is made more negative, the spectrum becomes
progressively denser and the levels descend toward the origin, the
lowest gap $\epsilon_2-\epsilon_1$ shrinking as the plateau narrows.
This is the entanglement-spectrum signature of the approach to the
gapless point reached as $J_t\to-\infty$.

\begin{figure}[htbp]
    \centering
    \includegraphics[width=\linewidth]{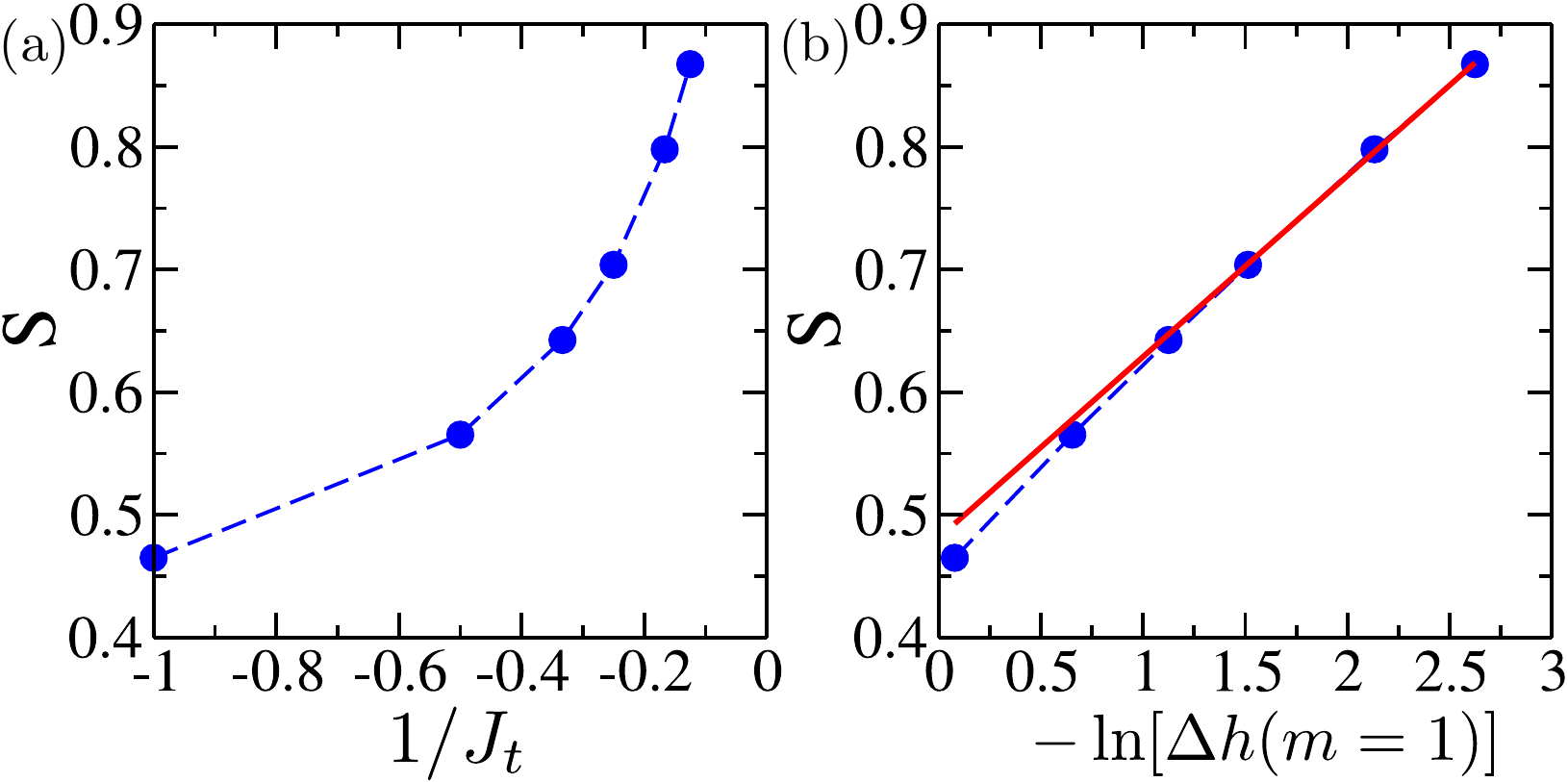}
    \caption{\label{fig:ent_entropy}Von Neumann entanglement entropy
    $S$ of the $m=1$ plateau, computed for a bipartition at the center
    of the open chain in the $S^z=L$ sector and extrapolated to
    $N\to\infty$ linearly in $1/L$. (a) $S$ as a function of $1/J_t$;
    the dashed line is a guide to the eye. The entropy grows with
    increasing concavity and does not saturate as $1/J_t\to0$.
    (b) $S$ versus $-\ln\Delta h(m=1)$, where $\Delta h(m=1)$ is the
    width of the $m=1$ plateau (Fig.~\ref{fig:hm1_width}); the red
    solid line is a linear fit, whose slope $\simeq0.15$ is close to
    $c/6=1/6$, the value expected for a $c=1$ Tomonaga-Luttinger
    liquid through $S=\frac{c}{6}\ln\xi+\mathrm{const}$ with
    $\xi\sim1/\Delta h(m=1)$.}
\end{figure}

We now turn to the entanglement entropy of the $m=1$ plateau, obtained
from the same Schmidt decomposition across the central
bond of the open chain in the $S^z=L$ sector. Splitting the chain into a
block $A$ and its complement, the von Neumann entanglement entropy is
\begin{equation}
S = -\,\mathrm{Tr}\,\rho_A\ln\rho_A = -\sum_i \lambda_i^2\ln\lambda_i^2,
\label{eq:vN}
\end{equation}
where $\rho_A$ is the reduced density matrix of block $A$ and
$\lambda_i$ are the Schmidt coefficients of the bipartition. For every
finite $J_t<0$, $S$ saturates as the system size is increased, the
area-law behavior expected of a gapped
plateau~\cite{eisert2010,laflorencie2016,schollwock2011}. Its saturated value, however,
grows steadily as the plateau narrows. In
Fig.~\ref{fig:ent_entropy}(a) we show $S$ as a function of $1/J_t$:
the entropy rises with increasing concavity and displays no sign of
saturation as $1/J_t\to0$. This is the entanglement counterpart of
the closing of the $m=1$ plateau reported in
Fig.~\ref{fig:hm1_width}. Indeed, as established in Sec.~\ref{sec:gs}
[Eq.~(\ref{eq:Heff})], in the $J_t\to-\infty$ limit the
$B$--$C$ pairs lock into triplets and the trimer chain maps onto a
spin-$1$ Heisenberg chain, with the $m=1$
plateau mapping onto the half-saturation point of that chain. The
spin-$1$ Heisenberg chain hosts no plateau at
half-saturation~\cite{oshikawa1997,sakai1991} and is instead a
critical Tomonaga-Luttinger liquid with central charge
$c=1$~\cite{giamarchi2004}.

At a critical, gapless point the entanglement entropy grows
logarithmically with the system size $L$, the only available length
scale; for open boundary conditions, with the entropy taken across a
single bipartition of the chain, $S=\frac{c}{6}\ln
L$~\cite{vidal2003,calabrese2004,calabrese2009,cheng2024a,ramos2026,schollwock2011}. Away from
criticality a finite correlation length $\xi$ cuts off this growth,
and for $L\gg\xi$ the entropy saturates at
$S=\frac{c}{6}\ln\xi+\mathrm{const}$; close to the transition
$\xi\sim1/\Delta$, set by the excitation gap $\Delta$. In the present
case the relevant gap is the
plateau width, $\Delta=\Delta h(m=1)$, which vanishes as
$J_t\to-\infty$ (Fig.~\ref{fig:hm1_width}); the correlation length
therefore diverges and $S$ grows without bound, as seen in
Fig.~\ref{fig:ent_entropy}(a). To make this quantitative, we plot $S$
against $-\ln\Delta h(m=1)\sim\ln\xi$ in
Fig.~\ref{fig:ent_entropy}(b). The data follow a straight line whose
slope, $\simeq0.15$ (i.e.\ $c\simeq0.9$), is close to the expected
$c/6=1/6$. The chain becomes strictly critical only at
$J_t\to-\infty$; at the finite couplings reached the gap $\Delta
h(m=1)$ is still nonzero, so the system is merely \emph{near} the
critical point, and the leading $\frac{c}{6}\ln\xi$ law receives
finite-gap corrections that account for the slope lying slightly below
$1/6$. The logarithmic
growth of $S$ thus provides an entanglement-based confirmation,
complementary to the vanishing of the plateau width, that the $m=1$
plateau closes onto the $c=1$ critical point of the spin-$1$ chain at
half-saturation.

\begin{figure}[htbp]
    \centering
    \includegraphics[width=0.9\linewidth]{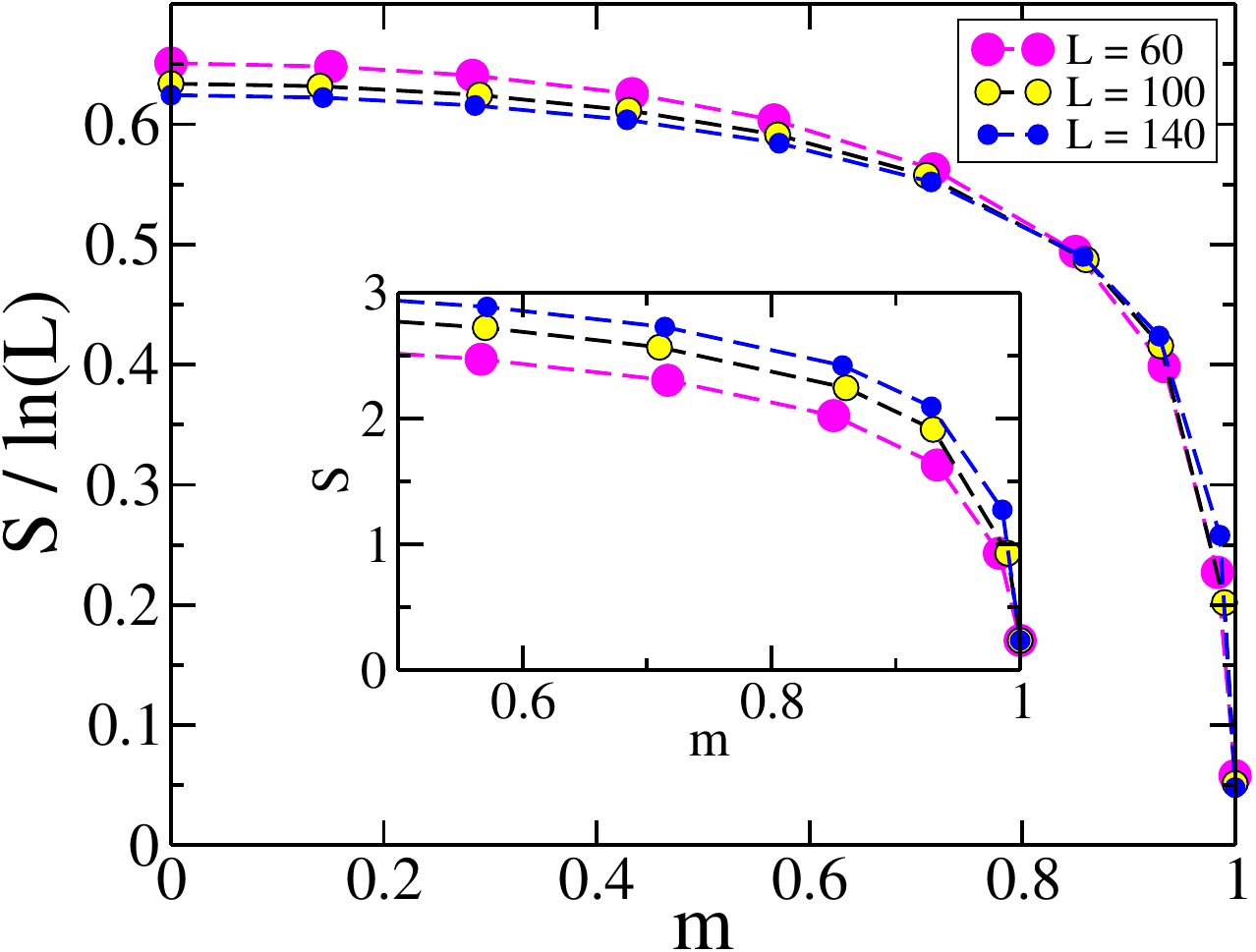}
    \caption{\label{fig:ent_lm}Main panel: von Neumann entanglement
    entropy $S$, divided by $\ln L$, as a
    function of the magnetization per unit cell $m=S^z/L$ in the
    Lieb-Mattis phase ($J_t=+1$), computed for a bipartition at the
    center of the open chain and system sizes $L=60$, $100$, and $140$
    unit cells. Each point corresponds to a member of the $S=L$
    ground-state multiplet. At $m=0$ the ratio $S/\ln L$ decreases with
    $L$ and extrapolates to $\simeq 0.60$, consistent with a
    logarithmic growth of the fixed-magnetization entropy of a spin
    multiplet~\cite{popkov2005}; at full polarization ($m=1$) it falls
    toward zero. Inset: the unnormalized entropy $S$ versus $m$ for the
    same sizes; at full polarization $S\simeq0.24$ is small and nearly
    size-independent, an almost product state across the cut.}
\end{figure}

It is instructive to contrast this behavior with the entanglement on
the Lieb-Mattis side ($J_t>0$), where the Lieb-Mattis theorem fixes the
ground state to the ferrimagnetic multiplet of total spin $S=L$. The
$m=1$ state is the fully aligned member $|S=L,S^z=L\rangle$, almost a
product across the central bond, with a small, size-independent
entanglement entropy $S\simeq0.24$ [inset of
Fig.~\ref{fig:ent_lm}]. The remaining members, reached by
lowering $S^z$, do not factorize under the cut: each half is itself a
ferrimagnet of definite magnetization, and since the total $S^z=mL$ is
fixed, the two are locked so that their magnetizations add up to $mL$.
The state is a superposition over the ways $mL$ is split between the
halves, and the entropy measures, logarithmically, how broadly this
splitting is distributed. The multiplet structure that enters is
therefore that of each half, not that of the chain as a whole.

Figure~\ref{fig:ent_lm} shows $S/\ln L$ as a function of $m$ in the
Lieb-Mattis phase ($J_t=+1$), for $L=60$, $100$, and $140$ unit cells.
The entropy is largest at $m=0$ and decreases monotonically to its
product-state value at full polarization. At $m=0$ the ratio $S/\ln L$
decreases with increasing $L$ and extrapolates to $\simeq 0.60$, which
signals that the entropy itself grows logarithmically with system size,
$S\simeq 0.60\,\ln L$. The same logarithmic growth occurs in an exactly
solvable case: for a fixed-magnetization member of the ground-state
multiplet of the Heisenberg ferromagnet, a central bipartition with
$L/2$ on each side gives $S=\tfrac{1}{2}\ln L+\mathrm{const}$~\cite{popkov2005},
the magnetization affecting only the additive constant. In both cases
the logarithm originates from the conserved magnetization being shared
between the two halves; the prefactor extracted here, $\simeq 0.60$, is
somewhat larger than the exact ferromagnetic value $\tfrac{1}{2}$. This
size dependence, and the overall variation of $S$ with $m$, are a direct
consequence of the SU(2) symmetry of the ferrimagnetic multiplet and
should not be read as bulk critical entanglement: unlike the
Haldane-side $m=1$ plateau, whose logarithmically growing entropy
signals the approach to the $c=1$ critical point, here it merely
reflects the multiplet structure of an otherwise gapped ferrimagnet. We note that the
intersublattice entanglement of an anisotropic ferrimagnetic spin
chain has recently been obtained within a spin-wave
approach~\cite{lee2025}, a setting distinct from the isotropic,
real-space bipartition considered here.

\section{Summary and discussions}
\label{sec:summary}

Using the density matrix renormalization group method, we
investigated the ground-state phases of a one-dimensional mixed-spin
$1$-$\tfrac{1}{2}$-$\tfrac{1}{2}$ trimer chain in which the spin-$1$
$A$ sites form a backbone with the spin-$\tfrac{1}{2}$ $B$ sites
through an antiferromagnetic exchange $J$, and the spin-$\tfrac{1}{2}$
$C$ sites are attached to the backbone as side groups through an
exchange $J_t$ that we allowed to take both signs. We computed
ground-state energies in each total-$S^z$ sector, magnetization
curves, plateau widths, site-resolved magnetizations, edge-state
profiles, and the entanglement spectrum and entropy.

The lattice is bipartite, with $A$ and $C$ sites on one sublattice
and $B$ sites on the other, and for $J_t>0$ all exchange couplings
are antiferromagnetic. The conditions of the Lieb-Mattis theorem
are thus satisfied, and the theorem fixes the ground-state
magnetization at $m=1$ per unit cell, in agreement with the robust ferrimagnetic plateau identified
in the magnetization curves and with the site-resolved
magnetizations, which display the characteristic
$(A+C)$ versus $B$ alternation. The plateaus at $m=0$, $m=1$, and at
the saturation value $m=2$ all comply with the
Oshikawa--Yamanaka--Affleck condition without spontaneous breaking
of translational symmetry. As $J_t$ is reduced and crosses zero, the
$C$ sites decouple and the backbone reduces to an alternating
spin-$(1,\tfrac{1}{2})$ Heisenberg chain, in agreement with the
crossing of the energy levels in different $S^z$ sectors observed in
the numerical data. Within this ferrimagnetic phase, the entanglement
entropy decreases monotonically along the ground-state multiplet, from
its largest value at $m=0$ to a small, size-independent value at full
polarization, reflecting how the conserved total magnetization, fixed
for the chain as a whole, can be shared between its two halves, each
itself a ferrimagnet of definite magnetization.

For $J_t<0$ the Lieb-Mattis premise of unfrustrated antiferromagnetic
exchange fails, and the
ground-state phases were therefore established directly from
the DMRG data. Two main features emerge: the $m=1$ plateau persists
but its internal structure reorganizes, with the $B$ sublattice
magnetization changing sign and the three sublattices $A$, $B$, and
$C$ becoming all polarized in the same direction as $|J_t|$ grows;
and a second plateau opens at $m=0$
and grows monotonically with $|J_t|$. In the strong-coupling limit
$J_t\to-\infty$, the $B$--$C$ pair locks into a triplet and the
trimer chain maps onto a uniform spin-$1$ Heisenberg chain with
effective coupling $J/2$, as captured by the effective Hamiltonian
of Eq.~(\ref{eq:Heff}). This mapping accounts both for the
extrapolated local magnetizations within the $m=1$ plateau and for
the extrapolated width of the $m=0$ plateau, $\Delta h(m=0)\simeq
0.196$, which matches half of the Haldane gap of the spin-$1$
Heisenberg chain. Consistently, the $m=1$ plateau closes as
$1/J_t\to 0$, since the spin-$1$ Heisenberg chain hosts no plateau
at half-saturation; the entanglement entropy of this plateau grows
logarithmically as it narrows, locating the closing at the $c=1$
Tomonaga-Luttinger critical point of that chain.

The topological character of the $m=0$ phase was established from
two complementary signatures. The local magnetization profile in
open chains exhibits exponentially localized boundary modes
compatible with fractional spin-$\tfrac{1}{2}$ edge states, with a
decay length that extrapolates to $\xi\simeq 5.7$ in the
$J_t\to-\infty$ limit, in close agreement with the correlation
length of the spin-$1$ Heisenberg chain. The entanglement spectrum
displays the even-fold degeneracy of its lowest levels for all
values of $J_t<0$ examined, confirming the symmetry-protected
topological order across the entire $m=0$ region. An AKLT-like
picture, in which the spin-$1$ at site $A$ decomposes into two
virtual spin-$\tfrac{1}{2}$ that lock into singlets with the real
spin-$\tfrac{1}{2}$ partners at $B$ and $C$, captures these features
qualitatively.

Our results provide a controlled characterization of how
Lieb-Mattis ferrimagnetism (for $J_t>0$) and Haldane-type
topological order (for $J_t<0$) are realized, in distinct sign
regimes of the side-coupling that meet at the decoupling point
$J_t=0$, in a quasi-one-dimensional mixed-spin geometry. Natural
extensions include the introduction of further-neighbor or
frustrating exchange paths, the investigation of similar branched
chains with different combinations of spin magnitudes, and the
search for material realizations of the
$1$-$\tfrac{1}{2}$-$\tfrac{1}{2}$ trimer topology in the family of
copper-based phosphates and related quasi-one-dimensional compounds.

\section*{Acknowledgements}
We acknowledge the support from Coordenação de Aperfeiçoamento de Pessoal de Nível Superior (CAPES), Grant No. 1575/2024, Conselho Nacional de Desenvolvimento Científico e Tecnológico (CNPq), through the Universal Call (Grant No.  407819/2024-0), and Fundação de Amparo à Ciência e Tecnologia do Estado de Pernambuco (FACEPE).

\section*{CRediT authorship contribution statement}
\textbf{A. Felinto:} Investigation, Software, Formal analysis, Data curation, Visualization, Writing -- review and editing. \textbf{R. R. Montenegro-Filho:} Conceptualization, Data curation, Funding acquisition, Investigation, Methodology, Software, Supervision, Writing -- original draft, Writing -- review and editing.

\section*{Declaration of competing interest}
The authors declare that they have no known competing financial interests or personal relationships that could have appeared to influence the work reported in this paper.

\section*{Data availability}
The data that support the findings of this study are available from the corresponding author upon reasonable request.

\section*{Declaration of generative AI and AI-assisted technologies in the manuscript preparation process}
During the preparation of this work the authors used Claude (Anthropic), models Claude Opus 4.7 and 4.8, in order to assist in drafting and editing the manuscript text and in developing data-analysis scripts. After using this tool/service, the authors reviewed and edited the content as needed and take full responsibility for the content of the published article.

\bibliographystyle{elsarticle-num}
\bibliography{TrimerosFelinto_els}
\end{document}